# Ultrafast Dynamics of Bilayer and Trilayer Nickelate Superconductors

Y. D. Li[1], Y. T. Cao[2,3], L. Y. Liu[1], P. Peng[1], H. Lin[1], C. Y. Pei[4], M. X. Zhang[4], H. Wu[5], X. Du[1], W. X. Zhao[1], K. Y. Zhai[1], J. K. Zhao[3], M.-L. Lin[5], P. H. Tan[5], Y. P. Qi[4,6,7], G. Li[4,6], H. J. Guo[3†], Luyi Yang[1,8†] and L. X. Yang[1,8†]

[1]*State Key Laboratory of Low Dimensional Quantum Physics, Department of Physics, Tsinghua University, Beijing 100084, China.*

[2]*Key Lab for Magnetism and Magnetic Materials of the Ministry of Education, Lanzhou University, Lanzhou 730000, China*

[3]*Songshan Lake Materials Laboratory, Dongguan, Guangdong 523808, China*

[4]*School of Physical Science and Technology, ShanghaiTech University and CAS-Shanghai Science Research Center, Shanghai 201210, China.*

[5]*State Key Laboratory of Superlattices and Microstructures, Institute of Semiconductors, Chinese Academy of Sciences, Beijing, 100083, People's Republic of China*

[6]*ShanghaiTech Laboratory for Topological Physics, Shanghai 200031, China.*

[7]*Laboratory of High-resolution Electron Microscopy, ShanghaiTech University, Shanghai 201210, China*

[8]*Frontier Science Center for Quantum Information, Beijing 100084, China.*

*Emails: LXY: lxyang@tsinghua.edu.cn; GL: ligang@shanghaitech.edu.cn; HJG: hjguo@sslab.org.cn; LYY: luyi-yang@mail.tsinghua.edu.cn;*

**In addition to the pressurized high-temperature superconductivity, bilayer and trilayer nickelate superconductors $La_{n+1}Ni_nO_{3n+1}$ (n = 2 and 3) exhibit many intriguing properties at ambient pressure, such as orbital-dependent electronic correlation, non-Fermi liquid behavior, and density-wave transitions. Here, using ultrafast reflectivity measurement, we observe a drastic difference between the ultrafast dynamics of the bilayer and trilayer nickelates at ambient pressure. Firstly, we observe a coherent phonon mode in $La_4Ni_3O_{10}$ involving the collective vibration of La, Ni, and O atoms, which is absent in $La_3Ni_2O_7$. Secondly, the temperature-dependent relaxation time diverges near the density-wave transition temperature of $La_4Ni_3O_{10}$, in drastic contrast to kink-like changes in $La_3Ni_2O_7$. Moreover, we estimate the electron-phonon coupling constants to be 0.05~0.07 and 0.12~0.16 for $La_3Ni_2O_7$ and $La_4Ni_3O_{10}$, respectively, suggesting a relatively minor role of electron-phonon coupling in the electronic properties of $La_{n+1}Ni_nO_{3n+1}$. Our work not only sheds light on the relevant microscopic interaction but also establishes a foundation for further studying the interplay between superconductivity and density-wave transitions in nickelate superconductors.**

Many unconventional superconductors exhibit a common interplay with density-wave states in their phase diagrams. The prime examples are the charge-density wave (CDW) and spin-density wave (SDW) in high-temperature cuprate superconductors [1-4] and iron-based superconductors [5-9]. Recently, high-temperature superconductivity has been discovered in layered nickelates in Ruddlesden-Popper (RP) phase under moderately high pressure [10]. Likewise, density-wave states have been revealed in the phase diagram of both bilayer and trilayer nickelates in RP-phase at ambient pressure [10-19]. Apparently, the exploration of the density-wave states in RP-phase nickelates will not only provide important insights into their normal state properties but also help construct a unified picture of unconventional superconductivity.

Ultrafast laser pulses possess a unique capability to disentangle different degrees of freedom at ultrashort time scale and extract dynamic information about ordered states, including relaxation process, microscopic coupling nature, and energy gap dynamics [20]. Based on the pump-probe technique, time-resolved probes have been widely used to investigate the density waves in different systems [21-24], including many nickelate compounds. In particular, ultrafast optical spectroscopy has revealed rich intriguing dynamic behaviors of nickelate superconductors, such as photo-excited phase fluctuations preserving long-range order [25], ultrafast charge localization and the dynamics of the pseudogap [26], vibrational symmetry breaking process [27], and the evidence for *d*-wave superconductivity [28].

In this work, by employing time-resolved reflectivity measurements, we systematically investigate the ultrafast dynamics of the bilayer and trilayer nickelates at ambient pressure. Despite their similar structure based on Ni-O layers, we observe drastic difference in their dynamic behaviors. The relaxation time of $La_4Ni_3O_{10}$ diverges around the CDW/SDW transition temperature, suggesting an energy gap of about $16.7 \pm 4.3$ meV, which is in drastic contrast to the weak kink-like features in $La_3Ni_2O_7$. The ultrashort laser pulse excites a coherent phonon mode involving collective motions of La, Ni, and O atoms, which is absent in $La_3Ni_2O_7$. More importantly, the extracted electron-phonon coupling (EPC) constant in $La_4Ni_3O_{10}$ is more than twice the value in $La_3Ni_2O_7$, consistent with the difference in the relaxation time of the two systems. Our investigation advances the understanding of the novel normal-state properties of newly discovered high-temperature nickelate superconductors, which sheds new light on the unconventional superconductivity in nickelates.

Figures 1a and 1b compare the crystal structures of $La_3Ni_2O_7$ and $La_4Ni_3O_{10}$ at ambient pressure. In each unit cell, two (three) Ni-O layers form a bilayer (trilayer) structure by sharing apical oxygen atoms. Temperature-dependent resistance measurements suggest a density-wave-like transition in $La_3Ni_2O_7$ as manifested by the kink-like features (Fig. 1c) [10, 15]. By contrast, a well-defined CDW/SDW transition is revealed by a rapid increase of the resistance near 132 K with decreasing temperature in $La_4Ni_3O_{10}$ (Fig. 1c), which is also evident by the cusp in the susceptibility and specific heat [29-32]. The resistance then peaks near 120 K and decreases monotonically down to the lowest temperature. Figure 1d compares the Raman spectra of $La_3Ni_2O_7$ and $La_4Ni_3O_{10}$ at room temperature. In the high-frequency range (200-600 cm$^{-1}$), the evident phonons locate near 564 cm$^{-1}$ (16.92 THz) and 396 cm$^{-1}$ (11.88 THz) in $La_3Ni_2O_7$ and $La_4Ni_3O_{10}$, respectively. Notably, $La_4Ni_3O_{10}$ exhibits more low-frequency phonons below 200 cm$^{-1}$, likely due to its more complicated trilayer structure.

Figure 2 shows the prototypical transient change of the optical reflectivity, $\Delta R/R =$

$[R(t) - R_0]/R_0$, where $R_0$ is the reflectivity measured before the pump pulse arrives. Overall, both the data of $La_3Ni_2O_7$ and $La_4Ni_3O_{10}$ show an instantaneous excitation followed by an exponential decay (Figs. 2a-b). By fitting the data at 80 K to a single-exponential decay (Supplementary notes 2 and 3), we obtain relaxation time around 3.68 ps and 0.57 ps for $La_3Ni_2O_7$ and $La_4Ni_3O_{10}$, respectively. The difference in their relaxation time suggests distinct relaxation processes in the two systems, consistent with their different resistance behaviors. Moreover, on top of the exponential decay, there is a periodic oscillation of the transient reflectivity of $La_4Ni_3O_{10}$ (inset of Fig. 2b), which is, however, absent in $La_3Ni_2O_7$.

The observation of the oscillation implies the excitation of coherent phonon modes, as evidenced by the Fourier transform of the oscillation data (Fig. 2c). With the help of density-functional-theory (DFT) calculation, the most prominent phonon mode at 3.87 THz (129 cm$^{-1}$) is identified as the $A_g$ mode involving collective motion of La, Ni, and O atoms (inset of Fig. 2c). Within the trilayer unit cell, the La atoms vibrate mainly in the plane. The outer Ni atoms exhibit both out-of-plane and in-plane vibration, while the inner Ni atoms stay put. The apical O atoms vibrate mainly in the plane, in contrast to the in-plane O atoms vibrating in both in-plane and out-of-plane directions (see the Supplementary note 7 and the Supplementary video for details). This phonon mode is also observed in the Raman data of $La_4Ni_3O_{10}$ but absent in $La_3Ni_2O_7$ (red arrow in Fig. 1d). Figure 2d summarizes the temperature evolution of the frequency of the coherent phonon (Supplementary note 1), which gradually softens with increasing temperature.

Such a collective motion with $A_g$ symmetry involving different vibrations of Ni-O atoms is, in principle, not possible in the bilayer $La_3Ni_2O_7$ due to the symmetry limitation. On the other hand, as inspired by the unique SDW state in $La_4Ni_3O_{10}$, where only the outer Ni-O layers show a periodic modulation of the spin moment [29], we propose that the excited coherent phonon with distinct motion of the three Ni-O layers may provide a coherent manipulation of the SDW order in $La_4Ni_3O_{10}$.

Figure 3 explores the temperature evolution of the ultrafast optical reflectivity. The false-color plots of the transient reflectivity of $La_3Ni_2O_7$ and $La_4Ni_3O_{10}$ in Figs. 3a and 3b reveal clear difference in their temperature-dependent behaviors. While $La_3Ni_2O_7$ exhibits a continuously prolonged relaxation time with the decreasing temperature, $La_4Ni_3O_{10}$ shows an abruptly increased relaxation time near the CDW/SDW transition temperature, in line with the previous observations in various density-wave materials [33-36]. The different temperature dependence of the dynamics in the two nickelates are further evident by the transient reflectivity change at selected temperatures shown in Figs. 3c and 3d.

To quantify the change of the dynamics across the CDW/SDW transition, we extract the amplitudes $A$ and relaxation times $\tau$ of the transient reflectivity change $\Delta R/R$ by fitting the data to a single-exponential decay (Supplementary notes 2 and 3). Figure 4 summarizes the results as functions of the temperature and the pump fluence. In $La_3Ni_2O_7$, both $A$ and $\tau$ decrease monotonically with the elevated temperature. We note the changes of the slope of $A$ and $\tau$ (Figs. 4a-b), which may be related to the kink-like features in the resistances and other transport measurements (black arrows in Fig. 1c) [29, 37]. Interestingly, we observe a linear temperature-dependence of the scattering rate $1/\tau$ as shown in the inset of Fig. 4b, which suggests a non-Fermi liquid behavior of the system, in nice agreement with the previous optical spectroscopy experiment [38].

The results in La$_4$Ni$_3$O$_{10}$ are qualitatively different from La$_3$Ni$_2$O$_7$. The amplitude shows a dip and the relaxation time shows a profound divergence near the CDW/SDW transition temperature (Figs. 4c, 4d, and Supplementary Note. 5). This divergent behavior has been widely observed in density-wave materials [33-36], which can be understood within the Rothwarf-Taylor (R-T) model aiming to elucidate the bottleneck effect caused by an energy gap [39]. The nice conformity between our data and the R-T model confirms the prototypical gap dynamics in the CDW/SDW state of La$_4$Ni$_3$O$_{10}$. By fitting the model to our data in Figs. 4c and 4d, we obtain an energy gap of $\Delta_0$ = 16.7 ± 4.3 meV (Supplementary note 5), in good agreement with the previous ARPES measurement [40]. It is worth noting that the reduced energy gap of La$_4$Ni$_3$O$_{10}$ $2\Delta_0/k_B T$ = 2.93 is less than the mean-field expectation.

We further estimate the EPC constant $\lambda$ based on the relation $\tau = \pi k_B T_e / 3\hbar\lambda\langle\omega^2\rangle$ [41-44], where $k_B$ and $\hbar$ are Boltzmann and reduced Planck constant, respectively, $\lambda\langle\omega^2\rangle$ is the second moment of the Eliashberg function, and $T_e$ is the electron temperature estimated with the two-temperature model [41, 42]. Using this simple model, we obtain $\lambda\langle\omega^2\rangle$ = 37.78 meV$^2$ and $\lambda\langle\omega^2\rangle$ = 123.55 meV$^2$ for La$_3$Ni$_2$O$_7$ and La$_4$Ni$_3$O$_{10}$, respectively (Supplementary note 6). After initial thermalization, the interaction with Debye phonons is the most efficient channel for the relaxation of hot carriers. It is therefore reasonable and commonly used to estimate $\langle\omega^2\rangle$ using Debye phonons, that is, $\langle\omega^2\rangle \propto \Theta_D^2$ [45-47], where $\Theta_D$ is the Debye temperature. With $\Theta_D$ = 383 K (33.0 meV) [48] and $\Theta_D$ = 459 K (39.6 meV) [12] for La$_3$Ni$_2$O$_7$ and La$_4$Ni$_3$O$_{10}$, we obtain $\lambda$ = 0.05~0.07 and $\lambda$ = 0.12~0.16, respectively (Supplementary Note 6). These values are consistent with the rough estimation using $\lambda \approx \hbar/\tau\Omega_D$ = 0.04 and 0.11 for La$_3$Ni$_2$O$_7$ and La$_4$Ni$_3$O$_{10}$, respectively, where $\Omega_D$ is the Debye frequency [49]. The EPC constant of La$_4$Ni$_3$O$_{10}$ agrees well with the value of $\lambda = 0.124$ given by the transport experiment [32] and those in other nickelates [50]. It is noteworthy that although the estimated EPC constant is too weak to support the observed high-temperature superconductivity, the role of EPC should not be completely excluded. Recent calculations have predicted an enhanced EPC constant as large as 1.75 in pressurized $La_3Ni_2O_7$ [51], which requires further experimental investigations.

Finally, we show the results measured under different pump fluences (*F*) at 80 K in Figs. 4e-h. Due to the increased photo-excited carriers, the amplitude of $\Delta R/R$ linearly increases with the pump fluence in $La_3Ni_2O_7$, with a signature of saturation or change of the slope at high-fluence regime (the damage threshold for $La_3Ni_2O_7$ crystals is about 100 μJ/cm$^2$) (Fig. 4e). Similar to the temperature-dependent experiments in Fig. 4b, the relaxation time continuously decreases with the pump fluence. In $La_4Ni_3O_{10}$, however, we observe a clear anomaly near 48 μJ/cm$^2$ in both the amplitude and relaxation time. In particular, the fluence-dependent relaxation time drastically increases at 48 μJ/cm$^2$, which fits nicely to the assumption that the relaxation time is inversely proportional to the gap $\Delta$ and the gap is suppressed by ultrashort laser pulse according to $\Delta = \Delta_0(1 - F/F_T)$ (red curve in Fig. 4h) [52], where $F_T$ is the threshold fluence required to suppress the density-wave order. The threshold fluence of 48 μJ/cm$^2$ corresponds to about 5.4 meV per unit cell, about one order larger than the electronic condensation energy of $N(E_F)\Delta^2/2 \approx 0.42$ meV, where $N(E_F)$ is the electronic density of states at $E_F$. This observation is on contrary to the results in the excitonic CDW material $TiSe_2$ [53] but in line with those in the prototypical Peierls CDW material $(K, Rb)_{0.3}MoO_3$, where a large amount of photon energy is deposited to the lattice [21, 54].

In summary, we present a comprehensive ultrafast optical spectroscopy study on $La_3Ni_2O_7$ and

La$_4$Ni$_3$O$_{10}$. While La$_4$Ni$_3$O$_{10}$ exhibits a prototypical density-wave dynamics, La$_3$Ni$_2$O$_7$ shows a weak transition-like behavior. Moreover, we observe the excitation of coherent phonon modes involving different vibrations of the three Ni-O layers in La$_4$Ni$_3$O$_{10}$, which may provide a route to coherently control the magnetism of the system. These observations, together with the larger EPC constant, suggest the lattice degree of freedom is more important in the electronic properties of La$_4$Ni$_3$O$_{10}$ compared to La$_3$Ni$_2$O$_7$. Our findings provide dynamic information on the interplay of multiple degrees of freedom in the nickelates, which will deepen our understanding of the normal state and superconducting properties of Ni-based high-temperature superconductors.

**Methods**

**Sample growth**

Single crystals of La$_3$Ni$_2$O$_7$ and La$_4$Ni$_3$O$_{10}$ were grown by the high-pressure optical floating zone technique [37]. Raw materials of La$_2$O$_3$ (99.99%) were heated at 900 ºC overnight to remove the undesired La(OH)3 phase. Subsequently, La$_2$O$_3$ and NiO (99.99%) powders were mixed in a stoichiometric ratio with 1% ~2% excess of NiO to compensate for its evaporation at high temperatures. The mixture was then pressed into pellets and sintered at 1100 °C in the air for 48 hours with several intermediate grindings. The reactant was pressed into rods with a diameter of about 6 mm and a length of about 14 cm under hydrostatic pressure and then sintered at 1300 °C for 2 hours. The single crystals of La$_3$Ni$_2$O$_7$ and La$_4$Ni$_3$O$_{10}$ were then grown in a high-pressure optical floating zone furnace (HKZ-300) with an oxygen pressure of 15 bar and 20 bar, respectively. The feed and seed rods were counter-rotated with speeds of 20 and 25 rpm, respectively. Growth rates of 2 mm/h and 5 mm/h were used for the $La_3Ni_2O_7$ and $La_4Ni_3O_{10}$ phases, respectively.

**Time-resolved reflectivity (Tr-reflectivity)**

Tr-reflectivity experiments were conducted using a Ti:sapphire oscillator with a center wavelength of 800 nm and a repetition rate of 80 MHz in a standard wavelength-degenerate pump-probe setup. Both the pump and probe beams are linearly polarized in a cross-polarization configuration to eliminate pump scattering. The pump and probe beams are focused to 26 μm and 12 μm on the sample, respectively. The intensity of the pump beam was modulated by an optical chopper at 3 kHz to facilitate lock-in detection. The overall temporal resolution is compressed to 40 fs using negative-dispersive mirrors. The reflectivity signal was detected by a balanced detector to mitigate laser power fluctuations. The sample was cleaved and subsequently kept in a cryostat under a vacuum of better than $1\times10^{-3}$ mbar during experiments.

**Ultralow-frequency Raman spectroscopy (ULF-Raman)**

Raman spectra at room temperature were acquired in the backscattering geometry with a Jobin–Yvon HR800 Raman system equipped with a 532 nm diode-pumped-solid-state laser, a 100× objective (NA = 0.9), and a liquid-nitrogen-cooled charge-coupled-device detector. Laser plasma lines were removed via BragGrate bandpass filters. High resolution of 0.32 cm$^{-1}$ was achieved by employing a grating with 2400 lines per mm. Three BragGrate notch filters were strategically applied to enable the lowest frequency down to 5 cm$^{-1}$. Temperature-dependent Raman data were similarly acquired, albeit with variations. A 785 nm solid-state laser and a

50× objective (NA = 0.55) were employed. An edge filter enabled the measurement down to 20 cm$^{-1}$ and a grating with 1200 lines per mm was used to achieve the spectral resolution of 0.36 cm$^{-1}$. The experimental setup incorporated a Montana cryostat system to achieve low temperatures and maintained the sample under a vacuum of 0.1 mTorr.

**Density-functional-theory calculations**

The density functional theory (DFT) calculations were performed using the Vienna Ab initio Simulation Package (VASP) [55, 56]. The electronic correlations were described by the Perdew-Burke-Ernzernhof functional [57]. The plane-wave cutoff energy was chosen to be 500 eV. Concerning the Brillouin zone integration, we use a 7 × 7 × 2 Monkhorst-Pack *k*-grid (spacing between *k*-points is 0.18 Å$^{-1}$). Structural relaxation was performed with a conjugated-gradient algorithm until the Hellmann-Feynman forces on each atom were less than 1 meV Å$^{-1}$ and the total energy was less than 10$^{-6}$ eV. To accelerate the calculation, we calculated the phonon dynamic matrix without expanding the primitive cell. The phonon spectrum and phonon eigenvectors were further obtained by using PHONOPY [58] with the frozen-phonon method.

**Acknowledgements**

This work is funded by the National Key R&D Program of China (Grant No. 2022YFA1403100, 2022YFA1403201, and 2022YFA1402703), the National Natural Science Foundation of China (No. 12274251, 12004270, 52272265), and the Guangdong Basic and Applied Basic ResearchFoundation (Grant No. 2022B1515120020). L.X.Y. acknowledges support from the Tsinghua University Initiative Scientific Research Program and the Fund of Science and Technology on Surface Physics and Chemistry Laboratory (No. XKFZ202102). G.L.

acknowledges the support of 2021-Fundamental Research Area 21JC1404700, and Sino-German Mobility program (M-0006).

**Author contributions**

L.X.Y. and L.Y. conceived the experiments. Y.D.L. carried out ultrafast reflectivity measurements with the assistance of L.Y.L, P.P., H.L., X.D., W.X.Z., K.Y.Z., L.Y.Y., and L.X.Y. *Ab-initio* calculations were performed by X.F.Z. and G.L. Single crystals were synthesized and characterized by Y.T.C., C.Y.P., J.K.Z., Y.P.Q., and H.J.G. Y.D.L. and H.W. carried out Raman measurements with the guidance of M.L.L. and P.H.T. The paper was written by Y.D.L., L.Y.Y., and L.X.Y. All authors contributed to the scientific planning, discussion, and manuscript revision.

**Competing interests**

Authors declare that they have no competing interests.

**Data and materials availability**

The data sets that support the findings of this study are available from the corresponding author upon reasonable request.

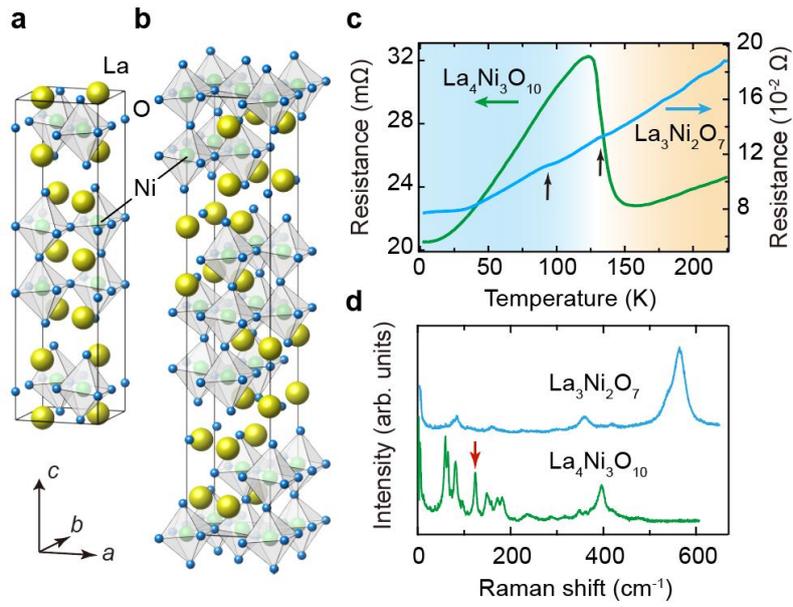

**Figure 1 | Basic properties of $La_3Ni_2O_7$ and $La_4Ni_3O_{10}$. a, b,** Crystal structures of $La_3Ni_2O_7$ (**a**) and $La_4Ni_3O_{10}$ (**b**). **c**, Temperature-dependent resistances of $La_3Ni_2O_7$ (blue curve) and $La_4Ni_3O_{10}$ (green curve) at ambient pressure, black arrows indicate kink-like features in $La_3Ni_2O_7$. **d**, 532 nm Raman spectra of $La_3Ni_2O_7$ and $La_4Ni_3O_{10}$, measured at room temperature.

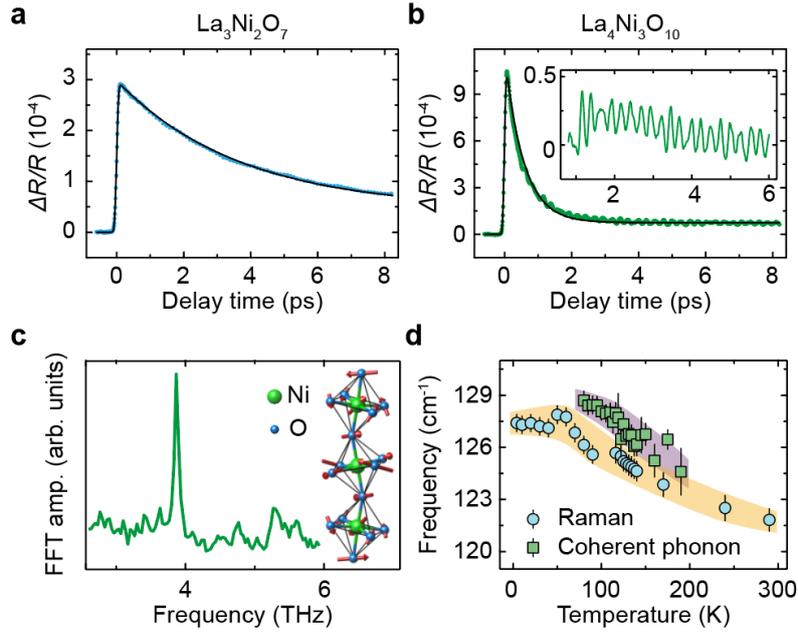

**Figure 2 | Excitation of coherent phonons in $La_4Ni_3O_{10}$. a**, Transient reflectivity change $\Delta R/R$ of $La_3Ni_2O_7$ at 80 K. **b**, $\Delta R/R$ of $La_4Ni_3O_{10}$ at 80 K showing the observation of coherent phonon vibrations. The inset shows the oscillatory part of the signal after subtracting the single-exponential background. **c**, Fourier transform of the oscillating part of the signal, the most prominent phonon mode locates at 3.87 THz. The inset shows the corresponding vibration of the atoms with the red arrows indicating the vibration directions. For simplicity, La atoms vibrating mainly in the plane are not shown (see the supplementary video). **d**, Temperature-dependence of the coherent phonon frequency of $La_4Ni_3O_{10}$ measured using ultrafast reflectivity and Raman experiments. Data of the ultrafast reflectivity were collected at the pump fluence $F = 7$ μJ/cm$^2$.

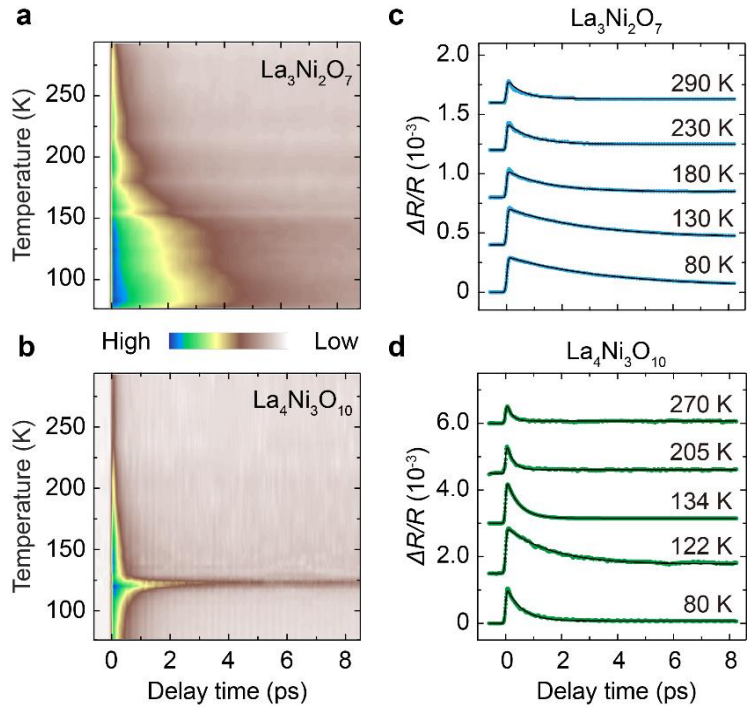

**Figure 3 | Temperature-dependent transient reflectivity change $\Delta R/R$. a**, **b**, The false-color plot of temperature-dependent transient reflectivity data of $La_3Ni_2O_7$ (**a**) and $La_4Ni_3O_{10}$ (**b**). **c**, **d**, Typical temporal evolution of $\Delta R/R$ in $La_3Ni_2O_7$ (**c**) and $La_4Ni_3O_{10}$ (**d**) at selected temperatures. The black curves are the phenomenological fit of the data. Data were collected at the pump fluence $F = 7\ \mu J/cm^2$.

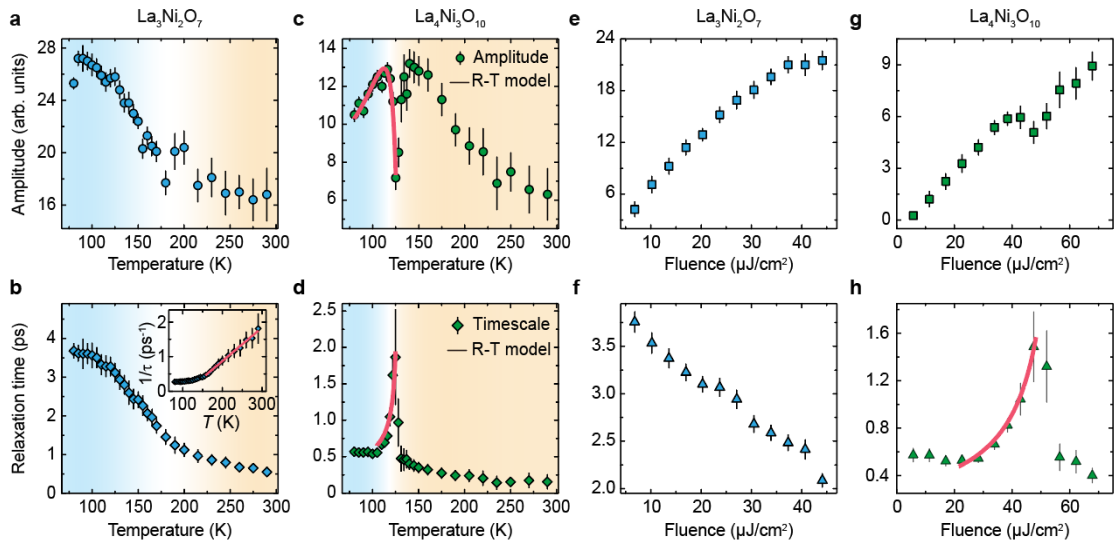

**Figure 4 | Evolution of the amplitude $A$ and relaxation time $\tau$ of transient reflectivity change. a**, **b,** Temperature dependence of $A$ (**a**) and $\tau$ (**b**) extracted in $La_3Ni_2O_7$. The inset of panel **b** shows the temperature dependence of $1/\tau$. **c, d**, The same as **a** and **b** but for $La_4Ni_3O_{10}$. The red curves are the fit of the data to the Rothwarf-Taylor (R-T) model. Data in **a-d** were collected at the pump fluence $F = 7$ μJ/cm². **e, f**, Fluence dependence of $A$ (**e**) and $\tau$ (**f**) extracted in $La_3Ni_2O_7$. **g, h**, The same as **e** and **f** but for $La_4Ni_3O_{10}$. The red curve in **h** is the fit of the data (see the main text). Data in **e-h** were measured at 80 K.